\documentclass[aps,pra,reprint,superscriptaddress]{revtex4-1}

\usepackage{lineno}

\usepackage{hyperref}
\usepackage{graphicx}  
\usepackage{dcolumn}   
\usepackage{bm}        
\usepackage{amssymb}   
\usepackage{epstopdf}
\usepackage{amsmath}
\usepackage{xspace}
\usepackage{verbatim}
\usepackage{color}

\usepackage{calrsfs}

\usepackage{soul}

\newcommand{\bra}[1]{\left\langle #1 \right|}
\newcommand{\ket}[1]{\left| #1 \right\rangle}

\newcommand{\up}{|\uparrow\rangle}
\newcommand{\updag}{\langle\uparrow|}
\newcommand{\down}{|\downarrow\rangle}
\newcommand{\downdag}{\langle\downarrow|}

\hypersetup{
    colorlinks=true,linkcolor=blue,citecolor=blue,
    filecolor=blue,urlcolor=blue,breaklinks=true
}

\begin{document}
\title{Ground-state cooling of a nanomechanical oscillator with N spins}

\author{V\'{i}ctor Montenegro}
\email{vmontenegro@fis.puc.cl}
\affiliation{Instituto de F\'{i}sica, Pontificia Universidad Cat\'{o}lica de Chile,
Casilla 306, Santiago, Chile}

\author{Ra\'{u}l Coto}
\affiliation{Instituto de F\'{i}sica, Pontificia Universidad Cat\'{o}lica de Chile,
Casilla 306, Santiago, Chile}
\affiliation{Universidad Mayor, Avenida Alonso de C\'ordova 5495, Las Condes, Santiago, Chile}

\author{Vitalie Eremeev}
\affiliation{Facultad de Ingenier\'{i}a y Ciencias, Universidad Diego Portales, Ej\'ercito 441, Santiago, Chile}

\author{Miguel Orszag}
\affiliation{Instituto de F\'{i}sica, Pontificia Universidad Cat\'{o}lica de Chile,
Casilla 306, Santiago, Chile}
\affiliation{Universidad Mayor, Avenida Alonso de C\'ordova 5495, Las Condes, Santiago, Chile}

\date{\today}

\begin{abstract}
Typical of modern quantum technologies employing nanomechanical oscillators is to demand few mechanical quantum excitations, for instance, to prolong coherence times of a particular task or, to engineer a specific non-classical state. For this reason, we devoted the present work to exhibit how to bring an initial thermalized nanomechanical oscillator near to its ground state. Particularly, we focus on extending the novel results of D. D. B. Rao \textit{et al.}, Phys. Rev. Lett. \textbf{117}, 077203 (2016), where a mechanical object can be heated up, squeezed, or cooled down near to its ground state through conditioned single-spin measurements. In our work, we study a similar iterative spin-mechanical system when $N$ spins interact with the mechanical oscillator. Here, we have also found that the postselection procedure acts as a discarding process, i.e., we steer the mechanics to the ground state by dynamically filtering its vibrational modes. We show that when considering symmetric collective spin postselection, the inclusion of $N$ spins into the quantum dynamics results highly beneficial. In particular, decreasing the total number of iterations to achieve the ground-state, with a success rate of probability comparable with the one obtained from the single-spin case.
\end{abstract}

\maketitle

\section{Introduction}

With the startling advancement of micro- and nano-fabricated quantum mechanical oscillators (NMO) \cite{OConnell2010-Ground, Teufel2011-GS, Chan2011, Peterson2016-GS, Yuan2015-GS}, the inclusion of high-quality nanomechanical devices for quantum technological purposes has become of pivotal importance. In the light of this, myriads of current quantum architectures have already implemented NMO as a central element to enable specific quantum tasks. For instance, in the domain of quantum sensing, ultra-sensitive measurement applications can be carried out by microscale force microscope cantilevers \cite{Treutlein2014, Knobel2003-Sensing}. Faithful conversion can be achieved from photonic states to the motion of a micrometer-sized mechanical resonator \cite{Reed2017}. And, in quantum networking schemes, NMO can serve as quantum transducer entailing two initial uncorrelated (or incompatible) subsystems \cite{Palomaki2013-Transfer, Stannigel2010-Transducer, Stannigel2011-Transducer}, to name a few.

Nonetheless, for many of the above implementations to hold, it is highly required to consider the NMO near to its ground state, or at least, counting with just a few quanta excitations on average (see also Refs. \cite{Palomaki2013-Transfer, Reed2017, Montenegro2017-MQS, Wang2012, McGee2013, Amir2011}). Thus, cooling methods of mechanical objects become transversely necessary to pave the way towards quantum technologies. Nowadays, the most common directions for ground-state cooling of NMO are \cite{Jaehne2008} passive optical cooling \cite{Schliesser2006, Wilson-Rae2004}, active optical feedback cooling \cite{Mancini1998, Arcizet2006, Kleckner2006, Martin2004}, and cooling via coupling to a different heavily damped solid state system \cite{MacQuarrie2017, Jaehne2008, Wilson-Rae2004, Rabl2010, Kepesidis2013} ---Recently, tripartite spin-optomechanical schemes have also been proposed for ground-state of NMO \cite{Restrepo2014, Zhou2016, Zeng2017, Nie15}. And, in quantum optomechanics in the non-resolved sideband regime \cite{Vanner2013}. The central idea behind several of the above routes consists in adding an extra dissipation channel, therefore, extracting energy from the mechanical object \cite{Liu2013, Marquardt2007, Ma2016, Gao2011, Stadler2014}, see also Ref. \cite{Abdi2017} for cooling of a two-dimensional mechanical membrane. Interestingly, a recent proposal of D. D. B. Rao \textit{et al.} \cite{Bhaktavatsala-2016} achieve full control of an NMO by coupling the mechanics to a single-spin state. The oscillator goes under a spin-induced thermal filter generated by conditioned spin measurements, where under certain circumstances the authors can heat-up, squeeze, and more particularly for our goal, to asymptotically cool down a nano-cantilever to its ground state. The process, analytically solved, it is probabilistic due to the iterative conditioned spin measurements required to obtain the desired state.

\begin{figure}
 \centering \includegraphics[width=\linewidth]{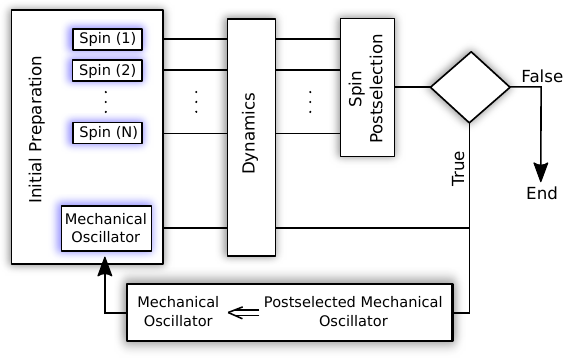}
 \caption{Schematic depiction of the iterative cooling process for the NMO. $N$ spins are coupled to a single-mode NMO. After the spin-mechanical system has evolved a time $t$, we proceed to postselect the $N$ spins. If the postselection results successful (True), we then repeat the process, where in the next step the NMO state will correspond to the one collapsed by the spin postselection.} \label{fig:model}
\end{figure}

In this paper, we present a theoretical proposal to cool down an NMO to its ground state through iterative spin postselections (see Fig. \ref{fig:model}). In particular, the spin-mechanical interaction is modeled via a conditioned displacement Hamiltonian, i.e., where the NMO displacement is conditioned on each spin eigenstate, thus resulting into an effective shifting of the NMO's potential center. The main idea of our work is early illustrated by considering an initial coherent mechanical state coupled to a quantum superposition of a single spin state. There, it is readily observed that a conditioned measurement of the spin subsystem (spin postselection) will collapse the NMO state into its ground state. We understand this process as a discarding step, in other words, we drive the NMO into its ground state by dynamically separating each mechanical state; the spin postselection step intends probabilistically to keep the mechanical mode as closer to the ground-state as possible. One may wonder whether some enhancement in the cooling rate (which is the figure of merit throughout our work) can be produced by coupling $N$-spins (independently) to the NMO and, at a later stage, optimally postselected. This last step accounts for a highly nontrivial task due to the large number of independent degrees of freedom $2^N - 1$. Although throughout this paper we study various spin postselections scenarios, one way to solve this issue is to consider a quite natural basis for this sort of dynamics, namely the collective spin angular momentum framework. There, for the symmetric case, we found that the inclusion of a collective spin postselection lead to a decreasing of the total number of iterations to achieve the ground-state with a non-negligible success probability.

This paper is organized as follows. In Section \ref{sec:motivation}, we motivate our work by solving the dynamics of a single-spin coupled to an NMO (initialized as a coherent state), we readily show that a conditioned spin measurement can drive the NMO into its ground state. In the following sections, we studied the role of the spin postselection when $N$ spins are interacting with the NMO (initialized as a thermal state). To address this case, we contrast three scenarios. Firstly, in Section \ref{subsection-a}, we proceed to postselect the spins independently. Here, it is shown that as $N$ increases fewer iterations must be performed to reach the NMO ground state. Secondly, in Section \ref{subsection-b}, we make use of spin postselection in correlated basis. Although this joint basis can drive the NMO to the desired state with fewer spins, this case suffers from a major drawback as the total success probability becomes rapidly negligible. Lastly, in Section \ref{subsection-c}, we show a quite balanced case, i.e. collective spin postselection, for which not only we can achieve the ground state faster, but also we can achieve it with a similar success probability when the spins are independently postselected. In Section \ref{sec:master-equation}, we study the feasibility of our scheme in an open quantum system. Concluding remarks are presented in Section \ref{sec:final-remarks}.

\section{Motivation : Cooling-down a coherent NMO with a single-spin}\label{sec:motivation}

For the sake of completeness, we will firstly illustrate the case where a single spin interacts with an NMO initialized in a coherent state. This simple dynamics will serve to build-up a most generic scenario, in which $N$ spins are independently coupled to the NMO and optimally postselected. In particular, we consider the following conditioned displacement Hamiltonian in a rotating frame at the spin frequency \cite{Montenegro-2014} ($\hbar = 1$)
\begin{equation}
 \hat{H}_\mathrm{int} = \hat{b}^\dag\hat{b} - \lambda \hat{\sigma}_{z}(\hat{b}^\dag + \hat{b})\label{eq:hint-single-spin},
\end{equation}

being $\lambda = x_\mathrm{zpf}\lambda_0/\omega_m$ the scaled (by the NMO frequency $\omega_m$) spin-mechanical coupling, $x_\mathrm{zpf}$ the NMO zero point amplitude and $\lambda_0$ the ``natural'' spin-mechanical coupling (in general, $\lambda_0$ depends on the geometry of the physical system) \cite{Treutlein2014}. As usual, $\hat{b}$ ($\hat{b}^\dagger$) is the boson annihilation (creation) operator for the single mode mechanical oscillator, whereas $\hat{\sigma}_{z}$ is the Pauli operator along the $z$-direction for the spin. The quantum dynamics can be calculated by considering the following unitary evolution operator (see Ref. \cite{Montenegro-2014} for details)
\begin{equation}
 \hat{U}(t) = \mathrm{exp}\left[\lambda \hat{\sigma}_{z}\left(\eta\hat{b}^\dagger - \eta^*\hat{b}\right)\right]\mathrm{exp}\left[-i\hat{b}^\dagger\hat{b}t\right],\label{eq:u-operator-sinle-spin}
\end{equation}

where $\eta \equiv 1 - e^{-it}$. To motivate our work, let us simply take the Hamiltonian in Eq. (\ref{eq:hint-single-spin}) together with an initial condition given by $1/\sqrt{2}(\up + \down)\otimes\ket{\beta}$ (with $\beta \in \mathbb{R}$ for simplicity). Thus, it is seen that each spin component will result into an effective center shift of the NMO's potential $\hat{b}^\dag\hat{b} \pm \lambda (\hat{b}^\dag + \hat{b})$, otherwise speaking, the NMO will follow two different oscillations according to the (unnormalized) wave-function
\begin{equation}
 \ket{\psi(t)} = \up\ket{\beta e^{-it} + \lambda\eta} + e^{-2i\lambda\beta\sin t}\down\ket{\beta e^{-it} - \lambda\eta}.
\end{equation}

From the above, let us suppose that we would like to decrease the phonon number accupation of the coherent mechanical object. To fulfill this task, we could simply collapse the above wave-function into one of its mechanical states by measuring, for instance, the $\up$ eigenstate (notice that the initial spin state could have been simply $\up$ instead of $1/\sqrt{2}[\up + \down]$. However, to link this initial spin state with the next section, we have preferred to keep the above spin superposition for clarity purposes) gives us
\begin{equation}
 \ket{\psi(t)}_m = \ket{\beta e^{-it} + \lambda\eta}.
\end{equation}

It is then straightforward to compute the NMO mean phonon number with the above postselected state as following
\begin{equation}
\langle \hat{b}^\dagger \hat{b}\rangle_\mathrm{post} = (\beta - \lambda)^2 + \lambda^2 + 2\lambda(\beta - \lambda)\cos t.
\end{equation}

To further illustrate this case, let us consider a spin postselection time to be $t = \pi$ (half of the NMO cycle). Hence, the ratio between the phonon accupation after spin postselection and the initial mean phonon number is reduced to
\begin{equation}
\frac{\langle \hat{b}^\dagger \hat{b}\rangle_\mathrm{post}}{\langle \hat{b}^\dagger \hat{b}\rangle_0} = \left(1 - 2\frac{\lambda}{\beta}\right)^2.
\end{equation}

For positive $\beta$ amplitude, the expression above can be found below unity for coupling values in the region $0 < \lambda < \beta$ (which is typically the case in several spin-mechanical systems). And more interestingly, the NMO can have zero phonons on average if the scaled coupling is chosen adequately as $\lambda = \beta/2$ --- a possible value for coherent amplitudes of $\beta \lesssim 2$.

At this point, we have presented a simple idea on how to decrease the phonons of a coherent NMO on average via optimal spin postselection. We wonder now whether we could extend this idea when the NMO is initially prepared as a thermal state, and also if there is any advantage on decreasing the NMO mean phonon number when $N$ spins are interacting with a thermalized oscillator. We devote the next section to cover all these questions.

\section{Cooling-down a single mode NMO with N-spins}\label{sec:cooling}

The following Section proposes to address if there is any advantage to cool down an NMO with $N$ spins. As shown in Section \ref{sec:motivation}, the critical stage of the scheme lies on the spin postselection step. Hence, if there is such an advantage, one may hypothesize that it should arise from a proper postselection of the $N$ spins. In what follows we studied three main cases in search for a solution, namely; i) each spin is to be measured independently, ii) $N$ spins are postselected in a joint basis, and iii) $N$ spins are collectively measured. The latter case being the most beneficial for the NMO cooling process.

\subsection{Postselecting the spins individually}\label{subsection-a}

Stimulated by the previous section, let us first consider a single spin directly coupled to a thermalized NMO (this case is also reported in Ref. \cite{Bhaktavatsala-2016}, here we would like to briefly recall the main results). Thus, we proceed to evolve the following initial state
\begin{equation}
 \hat{\rho}(0) = \ket{+}\bra{+}\otimes \frac{1}{\pi\bar{n}} \int \ket{\beta}\bra{\beta}e^{-\frac{|\beta|^2}{\bar{n}}}d^2\beta,\label{eq:initial-state-single-spin}
\end{equation}

being $\ket{+} = 1/\sqrt{2}(\up + \down)$, $\bar{n} = (\mathrm{exp}[\hbar\omega_m/k_BT) - 1])^{-1} \equiv \langle \hat{n} \rangle_0$ the thermal accupation phonon number, $T$ the distribution temperature, and $k_B$ the Boltzmann constant. Due to the election of the mechanical state in coherent basis, it is straightforward to write-down the density matrix for this case as \cite{Bhaktavatsala-2016, Montenegro-2014}
\begin{eqnarray}
 \nonumber &&\hat{\rho}(t) = \frac{1}{2\pi\bar{n}}\int d^2\beta e^\frac{-|\beta|^2}{\bar{n}} \Big(\up\updag \otimes \ket{\beta_\uparrow}\bra{\beta_\uparrow} + \\
 \nonumber && \down\downdag\otimes \ket{\beta_\downarrow}\bra{\beta_\downarrow} + \Big[\up\downdag\otimes \ket{\beta_\uparrow}\bra{\beta_\downarrow}e^{2i\vartheta} + h.c\Big] \Big)\\
 \label{eq:density-matrix-single-spin}
\end{eqnarray}

where $\vartheta \equiv \lambda r \cos(\phi - \frac{t}{2})\sin \frac{t}{2}$ and $\beta = r e^{i\phi}$. As discussed in the previous section, each qubit component displaces the mechanical state into $\beta_\uparrow = \beta e^{-it} + \lambda\eta$ or $\beta_\downarrow = \beta e^{-it} - \lambda\eta$. Nonetheless, contrary to the coherent case, here it is not readily accessible to know the effects produced by the spin non-diagonal terms in Eq. (\ref{eq:density-matrix-single-spin}). That being said, let us make two simple statements that actually can be done regarding the diagonal ones, namely i) the mechanical energy on average due to its dynamics alone is (i.e., by tracing out the spin degrees of freedom)
\begin{equation}
 \langle \hat{b}^\dagger \hat{b} \rangle = \bar{n} + 2\lambda^2(1 - \cos t),
\end{equation}

and ii), a projective spin measurement onto one of the eigenstates of $\hat{\sigma}_z$ will not attain a value below the initial phonon number accupation $\bar{n}$, i.e. both statements above show that mechanical cooling is not achieved under such circumstances. To investigate the role of the non-diagonal terms $\{\up\downdag,\down\updag\}$, let us postselect the density matrix shown in Eq. (\ref{eq:density-matrix-single-spin}) with a general vector parameterized in the Bloch sphere as
\begin{equation}
\ket{\psi}_\mathrm{target} = \cos(\theta/2)\up + \sin(\theta/2)e^{i\delta}\down.
\end{equation}

Hence, the normalized NMO state after the spin postselection reads as following
\begin{eqnarray}
\nonumber &\hat{\rho}(t)_\mathrm{post}& = \frac{1}{\mathcal{N}}\int d^2\beta \Big(\cos^2\frac{\theta}{2}\ket{\beta_\uparrow}\bra{\beta_\uparrow} + \sin^2\frac{\theta}{2} \ket{\beta_\downarrow}\bra{\beta_\downarrow} \\
&+& \Big[\cos\frac{\theta}{2}\sin\frac{\theta}{2}e^{i\delta} \ket{\beta_\uparrow}\bra{\beta_\downarrow}e^{2i\vartheta} + h.c\Big] \Big)e^\frac{-|\beta|^2}{\bar{n}},
\end{eqnarray}

with normalization constant
\begin{eqnarray}
 \mathcal{N} &=& \int d^2\beta e^\frac{-|\beta|^2}{\bar{n}} \Big(1 + \sin\theta\mathrm{Re}\Big[e^{i(\delta + 2\vartheta)}\bra{\beta_\downarrow}\beta_\uparrow\rangle\Big] \Big).
\end{eqnarray}

\begin{figure}
 \centering \includegraphics[width=\linewidth]{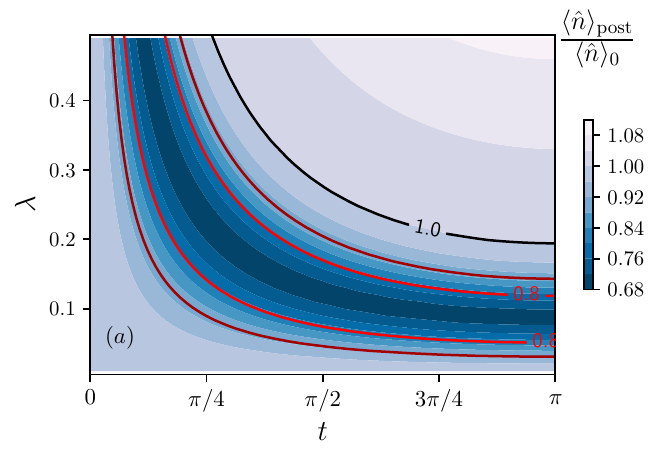}
 \centering \includegraphics[width=\linewidth]{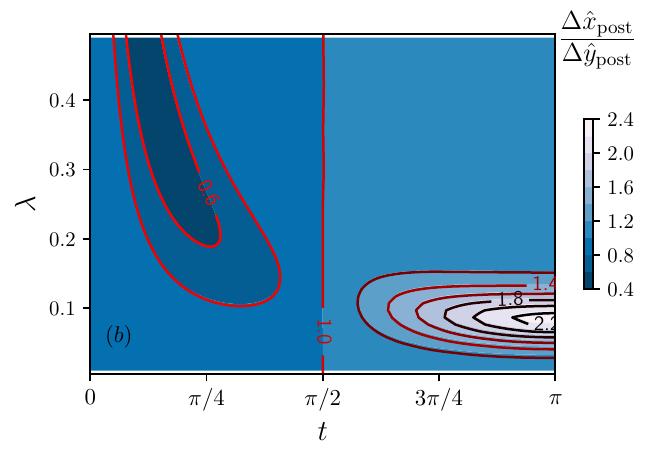} 
 \caption{Finding the parameters $\{t, \lambda\}$ to optimally cool-down the NMO. In (a), we show the ratio of the phonon number on average between the postselected state ($\langle \hat{n} \rangle_\mathrm{post}$) and the initial state ($\langle \hat{n} \rangle_0 = \bar{n}$). In (b) we illustrate the mechanical variances after spin postselection ($\Delta \hat{x}_\mathrm{post}/\Delta \hat{y}_\mathrm{post}$). Interestingly, only at $t = \pi/2$ both quadratures decrease simultaneously. Once $t = \pi/2$ is fixed, the optimal scaled coupling strength can be obtained from the top panel, being $\lambda \approx 0.12$.} \label{fig1:fig1}
\end{figure}

In the following, we will discuss the optimal parameters to achieve the mechanical cooling. For simplicity, after the spin-mechanical system has evolved a time $t$, we proceed to postselect the $\ket{+}$ spin state ($\theta = \pi/2$ and $\delta = 0$). In Fig. \ref{fig1:fig1}-a, we have plotted the ratio between the phonon number accupation with the postselected state ($\langle \hat{n} \rangle_\mathrm{post}$) and the initial energy of the NMO on average ($\langle \hat{n} \rangle_0$). As seen, there is a wide region of both time and coupling parameters where $\langle \hat{n}_\mathrm{post} \rangle < \langle \hat{n} \rangle_0$. Nevertheless, in Fig. \ref{fig1:fig1}-b we illustrate the ratio between the NMO variances, i.e. position (momentum) variance $\Delta \hat{x}_\mathrm{post}$ ($\Delta \hat{y}_\mathrm{post}$) when the spin has been postselected in $\ket{+}$. There, it is shown in fact that not every set of $\{t, \lambda\}$ values lead the mechanical state to cool-down, at least not in the sense of thermal cooling $\Delta \hat{x}_\mathrm{post} = \Delta \hat{y}_\mathrm{post}$. In Fig. \ref{fig1:fig1}-b, we can observe that when $t < \pi/2$ ($t > \pi/2$), the mechanical state is more likely to be position (momentum) squeezed at earlier (later) time. Consequently, only at a very specific time $t = \pi/2$ both quadratures decrease simultaneously $\Delta \hat{x}_\mathrm{post} \approx \Delta \hat{y}_\mathrm{post}$. Because of this, throughout our work we will uniquely consider the spin postselection time to be $t = \pi/2$ (even for $N$ spins interacting with an NMO), which also sets the optimal (dimensionless) spin-optomechanical strength $\lambda \approx 0.12$ (see Fig. \ref{fig1:fig1}-a). Notice, however, that for a single iteration the condition of $\Delta \hat{x}_\mathrm{post} \approx \Delta \hat{y}_\mathrm{post}$ holds for the usual (canonical) $x-$axis and $p-$axis in the phase space, where in reality the NMO state exhibits squeezing properties in a $\pi/2$-rotated ``new'' $x'-$axis and $p'-$axis (being the main axis of the squeezed state). Nonetheless, further iterations of the protocol assure that the NMO will be steered into a state with variances $\Delta \hat{x}_\mathrm{post} \approx \Delta \hat{y}_\mathrm{post}$ in the canonical axis, thus approaching to the desired ground-state cooling. With the above results, it is now clear that to cool-down an NMO from a thermal distribution, we must demand both to evolve the spin-mechanical dynamics from a spin superposition, as well as to postselect the spin subsystem onto a state with non-zero coherence. This, as opposed to the previous coherent case, where a spin postselection with zero coherence could drive the NMO to even zero phonons on average.

Additionally, from the top panel of Fig. \ref{fig1:fig1}, we can see that the best cooling rate that we could achieve is about $\langle \hat{n} \rangle_\mathrm{post} \approx 0.7 \langle \hat{n} \rangle_0$ with a single spin. One could ask, therefore, whether this ratio might be improved (i.e., $\langle \hat{n} \rangle_\mathrm{post} < 0.7 \langle \hat{n} \rangle_0$) with the inclusion of more spins coupled to the NMO. To investigate this case, we can generalize the Hamiltonian from a single-spin to $N$ independent spins interacting with a single mode NMO as following
\begin{equation}
 \hat{H}_\mathrm{int} = \hat{b}^\dag\hat{b} - \sum_{i=1}^N \lambda_i \hat{\sigma}_{z,i}(\hat{b}^\dag + \hat{b})\label{hint}.
\end{equation}

In the above Eq. (\ref{hint}), $\lambda_i = x_\mathrm{zpf}\lambda_0^{(i)}/\omega_m$ corresponds to the $i$-th scaled spin-mechanical coupling strength, and $\hat{\sigma}_{z,i}$ is the Pauli operator along the $z$-direction for the $i$-th spin.

Naturally, even though one could expect to assess a cooling enhancement when $N$ spins are considered in the dynamics (due to the $2^N$ mechanical displacements that will take place), the question regarding the optimal spin postselection becomes highly nontrivial. As a first result for the $N$ spin case, let us commence by evolving the system from an equiprobable spin superposition in conjunction with a thermal distribution for the NMO. Therefore, the initial state reads
\begin{equation}
\hat{\rho}(0) = \hat{\rho}_q(0) \otimes \frac{1}{\pi\bar{n}} \int \ket{\beta}\bra{\beta}e^{-\frac{|\beta|^2}{\bar{n}}}d^2\beta,
\end{equation}

where
\begin{equation}
\hat{\rho}_q(0) =  \frac{1}{2^N} \bigotimes_{i=1}^N \left(\up\updag + \up\downdag + \down\updag + \down\downdag\right)_i.\label{qubit0}
\end{equation}

It is straightforward to generalize the unitary evolution operator for this case, being 
\begin{eqnarray}
 \nonumber &&\hat{U}(t) = \mathrm{exp}\left[i \sum_{\{i,j\}=1}^N \lambda_i\lambda_j \hat{\sigma}_{z,i}\hat{\sigma}_{z,j} (t - \sin t)\right]\\
 &\times&\mathrm{exp}\left[\sum_{i=1}^N \lambda_i \hat{\sigma}_{z,i}\left(\eta\hat{b}^\dagger - \eta^*\hat{b}\right)\right]\mathrm{exp}\left[-i\hat{b}^\dagger\hat{b}t\right].\label{eq:u-operator}
\end{eqnarray}

Notice that for $N = 1$, the phase $\mathrm{exp}\left[i \sum_{\{i,j\}=1}^N \lambda_i\lambda_j \hat{\sigma}_{z,i}\hat{\sigma}_{z,j} (t - \sin t)\right]$ translates into a global phase, and hence we recover the unitary evolution operator shown in Eq. (\ref{eq:u-operator-sinle-spin}). Because the spins are both independent and linearly coupled to the NMO, we can notice that any spin arrangement are in fact interchangeable, i.e., the NMO will be displaced in the exact same amount if it is coupled to a spin state, for instance, $\up\down\down$, $\down\down\up$ or $\down\up\down$. Therefore, let us consider a generic array of $N$ spins given by $\ket{\{\uparrow\}_n,\{\downarrow\}_{N-n}} = \ket{\uparrow,\uparrow,\cdots,\downarrow,\downarrow,\cdots,\downarrow}$ (i.e., a vector state with $n$ spin-up ($\up$) and $N-n$ spin-down ($\down$) components) operated under the action of the unitary operator (\ref{eq:u-operator}),
\begin{eqnarray}
 \nonumber &&\hat{U}(t)\ket{\{\uparrow\}_n,\{\downarrow\}_{N-n}} = e^{4i\lambda^2[n(n-N) - m(m-N)](t - \sin t)}\\
 &\times&\hat{D}[(2n-N)\lambda\eta]e^{-i\hat{b}^\dagger\hat{b}t}\ket{\{\uparrow\}_n,\{\downarrow\}_{N-n}}, \label{eq:u-mechanics}
\end{eqnarray}

being $\hat{D}[\cdots]$ the mechanical displacement operator, and where for simplicity we have considered $\lambda_i = \lambda_j = \lambda$. Therefore, a coherent state for the mechanical object will be displaced in a quantity
\begin{equation}
 \hat{D}[(2n-N)\lambda\eta]\hat{D}[\beta e^{-it}]\ket{0} = e^{i\theta}\ket{\beta e^{-it} + \lambda(2n-N)\eta} \label{eq:displacement}
\end{equation}

where we recall that we have considered complex coherent amplitude $\beta = r e^{i\phi}$, and thus $\theta = 2 \lambda (2n-N) r\cos(\phi - t/2)\sin(t/2)$ was calculated by using the relationship $\hat{D}[\alpha_1]\hat{D}[\alpha_2] = e^{(\alpha_1\alpha_2^* - \alpha_1^*\alpha_2)/2}\hat{D}[\alpha_1 + \alpha_2]$.

With the aid of Eqs. (\ref{eq:u-mechanics}) and (\ref{eq:displacement}), and post-selecting the spins' degrees of freedom with a target state
\begin{equation}
 \ket{\psi}_\mathrm{target} = \frac{1}{2^{\frac{N}{2}}} \bigotimes_{i=1}^N (\up + \down)_i, \label{eq:ket-+-i}
\end{equation}

we can write-down the postselected state for the mechanics as following:
\begin{eqnarray}
 \nonumber \hat{\rho}(t)_\mathrm{post} &=& \frac{1}{2^{2N}\pi\bar{n}\mathcal{P}}\sum_{n,m=0}^N \mathcal{C}_{n,m} \times\\
 &&\int d^2\beta e^{i (\theta_n - \theta_m)}e^{-\frac{|\beta|^2}{\bar{n}}}\ket{\varphi_n(t)}\bra{\varphi_m(t)},\label{eq:full-density-matrix-N}
 \end{eqnarray}
 
where we have defined
\begin{eqnarray}
 \nonumber \mathcal{C}_{n,m} &=& \binom{N}{N-n}\binom{N}{N-m}e^{4i(t - \sin t)[n(n-N) - m(m-N)]},\\
 \nonumber \varphi_n(t) &=& \beta e^{-i t} + \lambda(2n - N)\eta,\\
 \nonumber \theta_n &=& 2\lambda(2n - N)r\cos(\phi - t/2)\sin(t/2),\\
 \nonumber \mathcal{P} &=& \frac{1}{2^{2N}\pi\bar{n}} \sum_{n,m=0}^N \mathcal{C}_{n,m} \times\\
 &&\int d^2\beta e^{i (\theta_n - \theta_m)}e^{-\frac{|\beta|^2}{\bar{n}}}\bra{\varphi_m(t)}\varphi_n(t)\rangle.
\end{eqnarray}

\begin{figure}
 \centering \includegraphics[width=\linewidth]{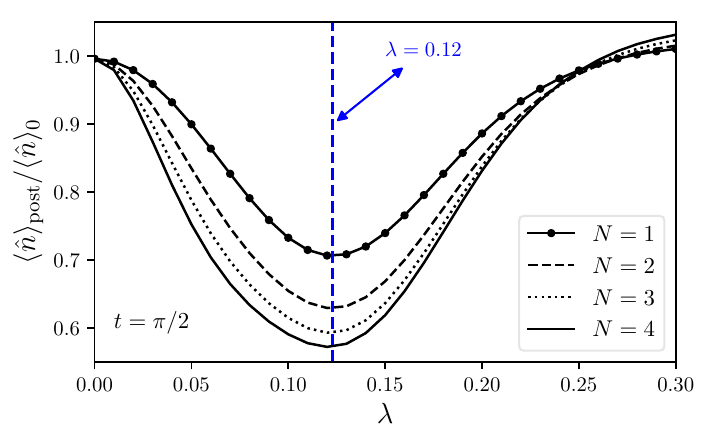}
 \caption{Ratio of the phonon number on average ($\langle \hat{n} \rangle_\mathrm{post}/\langle \hat{n} \rangle_0$) as a function of the coupling parameter ($\lambda$) for several spins coupled to the NMO, where $N$ stands for the number of the interacting spins. We have considered the spin postselection as in Eq. (\ref{eq:ket-+-i}).} \label{fig2:fig2}
\end{figure}

We now have explicitly derived the density matrix for the NMO when $N$ spins are independently postselected as shown in Eq. (\ref{eq:ket-+-i}). As a next step, we calculate in Fig. \ref{fig2:fig2}, the mean phonon energy ratio ($\langle \hat{n} \rangle_\mathrm{post}/\langle \hat{n} \rangle_0$) versus the coupling value ($\lambda$) for a spin-mechanical system with up to four spins ($N=4$) interacting with the NMO. To calculate $\langle \hat{n} \rangle_\mathrm{post} = \mathrm{Tr}[\hat{n}\hat{\rho}(t)_\mathrm{post}]$, we have used the analytical expression shown in Eq. (\ref{eq:full-density-matrix-N}), where for the numerics we considered $t=\pi/2$ and $\bar{n} = 10$. From Fig. \ref{fig2:fig2} we can readily notice two results, namely i) the optimal coupling value $\lambda \approx 0.12$ for which the ratio reaches its minimal value does not depend on the numbers of postselected spins ($N$). And ii) although increasing $N$ improves the mechanical cooling effect, the fact that the enhancement ratio $\langle \hat{n} \rangle_\mathrm{post}^{(N+1)} / \langle \hat{n} \rangle_\mathrm{post}^{(N)}$ goes rapidly towards unity, it makes considering $N>4$ redundant. For instance, for $N = 5$ then $\langle \hat{n} \rangle_\mathrm{post}^{(6)} / \langle \hat{n} \rangle_\mathrm{post}^{(5)} \approx 0.98$.

As a consequence of the above nonlinear enhancement ratio, it is unviable to fulfill the task of cooling-down an NMO to its ground state by uniquely postselecting $N$ spins [as in Eq. (\ref{eq:ket-+-i})] only once. Because of this, we proceed to iterate the protocol as follows: once the spin-mechanical system has evolved a time $t = \pi/2$, we perform a spin postselection collapsing the mechanical state as shown in Eq. (\ref{eq:full-density-matrix-N}). Naturally, if the postselection happens to be successful the spin state will also be found in the state $\ket{\psi}_\mathrm{target} = 1/(2^{\frac{N}{2}}) \bigotimes_{i=1}^N (\up + \down)_i$, for which therefore we continue to repeat the above steps.

\begin{figure}
 \centering \includegraphics[width=\linewidth]{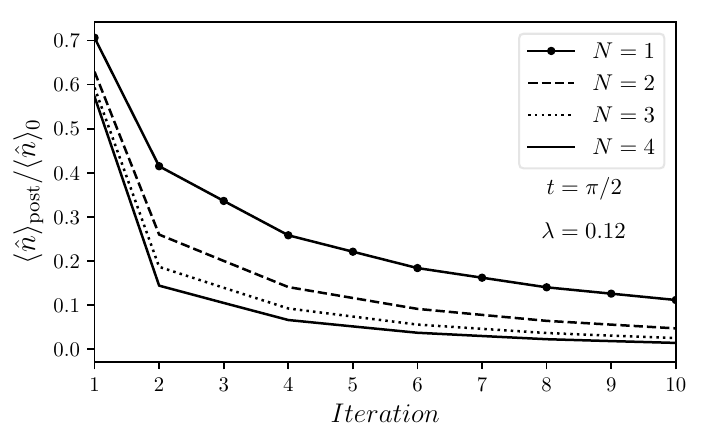}
 \centering \includegraphics[width=\linewidth]{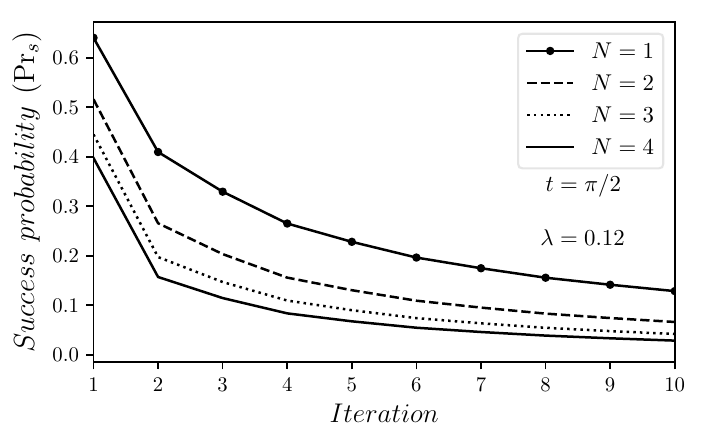}
 \caption{In the top panel, we show the ratio of the phonon number on average ($\langle \hat{n} \rangle_\mathrm{post}/\langle \hat{n} \rangle_0$) against the number of iterations for several spins coupled to the NMO. In the bottom panel, we illustrate the success probability at each iteration.} \label{fig3:fig3}
\end{figure}

Numerical results for the iteration procedure \cite{Johansson2013} are shown in Fig. \ref{fig3:fig3}-a. There, we have considered the optimal values $t=\pi/2$ and $\lambda = 0.12$ assuring ``thermal'' cooling. More importantly, controlling $N=4$ spins during the whole iteration process result in a considerable reduction of iterations. Moreover, in Fig. \ref{fig3:fig3}-b, we have plotted the success probability showing that we could achieve the similar cooling outcome regarding both probability and its cooling ratio $\langle \hat{n} \rangle_\mathrm{post}/\langle \hat{n} \rangle_0$ in less iterations. For instance, the cooling procedure with a single spin iterated eight times (hence eight postselections must be performed) can be efficiently converted by the one with four spins iterated only twice (although a same total number of postselections must be realized).

\subsection{Spin postselection with correlated basis}\label{subsection-b}

In the previous section, we have investigated the case where $N \geq 1$ spins are coupled to an NMO and, particularly, we have realized an independent postselection on each spin to decrease the mechanical energy on average. Nonetheless, when $N > 1$ the optimal postselection becomes highly nontrivial, mainly due to the $2^N - 1$ free parameters to be considered in the optimization procedure. In this Section, we will study the case where the postselection is carried out using correlated basis. The simplest nontrivial case is to consider $N=2$, where for our purpose we will make use of the following well-known correlated basis
\begin{eqnarray}
 \ket{\Phi^-} &=& \frac{1}{\sqrt{2}}(\down\down - \up\up),\\
 \ket{\Phi^+} &=& \frac{1}{\sqrt{2}}(\down\down + \up\up),\\
 \ket{\Psi^-} &=& \frac{1}{\sqrt{2}}(\down\up - \up\down),\\
 \ket{\Psi^+} &=& \frac{1}{\sqrt{2}}(\down\up + \up\down). 
\end{eqnarray}

Above, the four Bell vectors $\{\ket{\Phi^-}, \ket{\Phi^+}, \ket{\Psi^-}, \ket{\Psi^+}\}$ form a $2 \times  2$ basis, i.e. any vector spanned in the computational basis (with $\sum_i |c_i|^2 = 1$)
\begin{equation}
 \ket{\Theta} = c_1\up\up + c_2\up\down + c_3\down\up + c_4\down\down 
\end{equation}

can be written in terms of the Bell basis as following:
\begin{eqnarray}
 \ket{\Theta} &=& \frac{1}{\sqrt{2}}\Big[ (c_4 + c_1)\ket{\Phi^+} + (c_4 - c_1)\ket{\Phi^-}\\
 &+& (c_3 + c_2)\ket{\Psi^+} + (c_3 - c_2)\ket{\Psi^-} \Big].
\end{eqnarray}

Certainly, the cooling scheme remains the same, only the target spin for the postselection has now changed to be performed in a joint basis 
\begin{eqnarray}
\ket{\psi}^{(2)}_\mathrm{target} &=& b_1\ket{\Phi^-} + b_2\ket{\Phi^+} + b_3\ket{\Psi^+} + b_4\ket{\Psi^-},\\
1 &=& |b_1|^2 + |b_2|^2 + |b_3|^2 + |b_4|^2.
\end{eqnarray}

Numerical simulations (with Real coefficients) show that the optimal target spin is
\begin{eqnarray}
 \ket{\psi}^{(2)}_\mathrm{target} &=& \frac{1}{2}(\down\down + \up\up) + \frac{1}{\sqrt{2}}\down\up,\\
 &=& \frac{1}{\sqrt{2}} \ket{\Phi^+} + \frac{1}{2}(\ket{\Psi^+} + \ket{\Psi^-}).
\end{eqnarray}

Furthermore, another simulation run, where we have considered up to three spins, gives us the following optimal correlated target spin
\begin{eqnarray}
 \nonumber \ket{\psi}^{(3)}_\mathrm{target} &=& a( \up\up\up + \up\up\down + \up\down\up \\
 \nonumber &+&  \up\down\down + \down\down\up + \down\down\down ) \\
 &+& \frac{1}{5}(\down\up\up + \down\up\down),\\
 a &=& -\sqrt{\frac{1}{6}\left(1 - \frac{2}{25}\right)}. \label{eq:target-with-3-correlated}
\end{eqnarray}

\begin{figure}
 \centering \includegraphics[width=\linewidth]{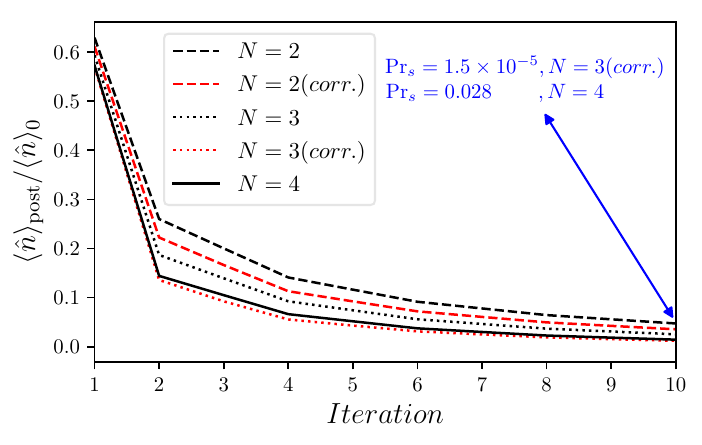}
 \caption{We compare the ratio of the phonon number accupation at each iteration when $N=2$, $N=3$ and $N=4$ spins are postselected independently, with the case when two and three spins are postselected in a joint basis, $N=2(corr.)$ and $N=3(corr.)$, respectively. We also show that, although the ratio $\langle \hat{n} \rangle_\mathrm{post} / \langle \hat{n} \rangle_0$ for $N=3(corr.)$ can in fact surpass $N = 4$, its negligible success probability ($\mathrm{Pr}_s$) at the tenth iteration makes it impracticable. Other values are $t = \pi/2$, $\lambda = 0.12$, and $\langle \hat{n} \rangle_0 = 10$.} \label{fig6:fig6}
\end{figure}

In Fig. \ref{fig6:fig6} we compare the ratio of the phonon accupation number from the previous section (where we have postselected the spins independently) with the spin postselection making use of correlated spin basis. Notice that, in this present case, after the spins are postselected, we require to restart the spins in the subsequent iteration such as in Eq. (\ref{qubit0}). This could be considered as a disadvantage on itself compared with the former case, where no reinitialization of the spin subsystem is necessary. On the other hand, as seen in Fig. \ref{fig6:fig6}, and considering the optimal target spin state with $N = 3$ [Eq. (\ref{eq:target-with-3-correlated})], the result is translated into a slight enhancement of the cooling ratio when contrasted to the case of postselecting $N = 4$ spins independently. Concerning probability, we have embedded in Fig. \ref{fig6:fig6} the success probability for the 10th-iteration. Certainly, for the case of $N=3$ with correlated basis it is seen to be extremely unfeasible, as the success probability turns out to be of the order of $\sim 10^{-5}$ (and situated way below the case of $N=4$ with a success probability of $0.028$). Moreover, even for the sixth iteration using a joint basis its success probability is quite modest, being a value of $\approx 0.006$ ---still far below the independent $N=4$ case at the tenth iteration.

The above analysis shows two notable disadvantages when postselecting $N$ spins utilizing correlated basis. Firstly, the optimization of the target spin is numerically implausible. Even for a few numbers of spins coupled to the NMO such as $N = 4$, the numerics takes larger times to be fully optimized. Secondly, and more important, even if the target is optimized, the usage of correlated basis is not comparable to the independent spin postselection case because of the very low success probabilities. In the above study, we have found that although the cooling rate could be improved, the scheme suffers from not being genuinely feasible. The next section aims to find a case where the cooling rate is improved at each iteration by almost no cost of the success probability. We will show that this is indeed the case when considering the collective configuration of the spins.

\subsection{Collective spin postselection}\label{subsection-c}

At this very stage of our work, we have addressed the system by considering either independent spin postselection or measurement on a joint basis for the spin subsystem, i.e., $2^N$ degrees of freedom. It has been shown in the latter case that, under specific optimal joint measurements the mean phonon number decreases faster than in the independent measurement case. However, finding the optimal target state is far from being a straightforward computational task (even for just a few spins coupled to the NMO, such as $N=4$). In fact, the independent spin degrees of freedom ($2^N - 1$) results in a highly demanding way to achieve the desired mechanical ground-state cooling. To surpass this difficulty, we proceed to describe the spin-mechanical evolution in a very natural collective spins manner. This description has been extensively studied in the cavity-QED regime, more particularly in the so-called Dicke model (where homogenous coupling between the emitters and the quantized field mode is usually considered. Inhomogeneous Dicke model has also been studied, see Ref. \cite{Tsy2010}), where $N$ independent emitters interact with a single mode of the quantized electromagnetic field. Typically, in the Dicke model, the collective operators are introduced as following 
\begin{eqnarray}
\hat{S}_z &=& \sum_{i=1}^N \hat{\sigma}_{z,i},\\
\hat{S}_\pm &=& \sum_{i=1}^N \hat{\sigma}_i^{\pm},
\end{eqnarray}

satisfying the angular momentum algebra
\begin{eqnarray}
\hat{S}_z|s,m,\mathcal{D}\rangle &=& m|s,m,\mathcal{D}\rangle, \hspace{1cm} -s \leq m \leq +s \label{eq:collective-z}\\
\hat{S}^2|s,m,\mathcal{D}\rangle &=& s(s+1)|s,m,\mathcal{D}\rangle, \hspace{1cm} \hat{S}^2 = \textbf{S} \cdot \textbf{S},\\
\hat{S}_\pm\ket{s,m} &=& \sqrt{(s \mp m)(s \pm m + 1)}\ket{s, m \pm 1}.
\end{eqnarray}

In the above, $m$ is the angular momentum's component along the $z$ direction ($|m| \leq s$), $s$ is related to the total angular momentum ($s \leq N/2$, $s_{min} = 0, 1/2$ for $N$ as an even or odd number, respectively), and $\mathcal{D}$ stands for the degeneracy parameter (with respect to the irreducible representation of the uncoupled spin basis); each Dicke state $\ket{s,m, \mathcal{D}}$ has a degeneracy \cite{Garraway2011}

\begin{equation}
 \mathcal{D}_{N,s} = (2s + 1) \frac{N!}{\left( \frac{N}{2} + s + 1 \right) !\left( \frac{N}{2} - s \right)!}.
 \end{equation}

As seen from the above, the states with $s = N/2$ have no degeneracy and are symmetric.

The unitary evolution operator for the Hamiltonian in collective representation

\begin{equation}
\hat{H} = \hat{b}^\dagger \hat{b} - \lambda \hat{S}_z (\hat{b}^ \dagger + \hat{b})
\end{equation}

reads as following
\begin{equation}
\hat{U}(t) = e^{i\lambda^2 \hat{S}_z^2 (t - \sin t)}e^{i\lambda \hat{S}_z (\eta \hat{b}^\dagger - \eta^* \hat{b})}e^{-i \hat{b}^\dagger \hat{b} t}.
\end{equation}

We now proceed to consider the initial condition as an array of non interacting spins coupled to a thermalized NMO (we have dropped the $\mathcal{D}$ degeneracy parameter, as we are only considering the symmetric case, $s = N/2$)
\begin{equation}
\hat{\rho}(0) = \sum_{\substack{m=-s,s\\ m'=-s',s'}} c_{m}c_{m'}|s,m\rangle \langle s',m'|\otimes \hat{\rho}(0)_\mathrm{NMO}.
\end{equation}

Therefore, the dynamics can easily be written as
\begin{eqnarray}
\nonumber \hat{\rho}(t) &=& \sum_{\substack{m=-s,s\\ m'=-s',s'}} e^{i\lambda^2(t - \sin t)(m^2 - m'^2)}c_{m}c_{m'}|s,m\rangle \langle s',m'|\\
&\otimes& \hat{D}[\lambda m \eta] e^{-i \hat{b}^\dagger \hat{b} t} \hat{\rho}(0)_\mathrm{NMO} e^{+i \hat{b}^\dagger \hat{b} t} \hat{D}[\lambda m' \eta]^\dagger.
\end{eqnarray}

Following the same procedure as before, we now proceed to postselect the spins with the target state
\begin{equation}
|\psi\rangle_{\mathrm{target}} = \sum_{m''=-s''}^{s''} d_{m''}|s'',m''\rangle.
\end{equation}

Thus, the unnormalized state is:
\begin{eqnarray}
\nonumber \hat{\rho}(t) &=& \sum_{\substack{m=-s,s\\ m'=-s',s'}} e^{i\lambda^2(t - \sin t)(m^2 - m'^2)}c_{m}c_{m'} d_{m}d_{m'}\\
&\otimes& \hat{D}[\lambda m \eta] e^{-i \hat{b}^\dagger \hat{b} t} \hat{\rho}(0)_\mathrm{NMO} e^{+i \hat{b}^\dagger \hat{b} t} \hat{D}[\lambda m' \eta]^\dagger.
\end{eqnarray}

Certainly, when considering $s = N/2$ the dimension of the spin subsystem has been reduced linearly from $2^N$ to $2s + 1 = N+1$. Moreover, the collective behavior could be exploited by an appropriate election of the $c_i$ and $d_j$ independent parameters, being the preselection and postselection weights, respectively. In principle, those weights can be prepared in any distribution. Nonetheless, for the sake of simplicity we have determined to consider firstly only flat distributions of the above, i.e., $c_i = d_i = \sqrt{1/(N+1)}, \forall i \in \mathbb Z$. Other distributions, such as Gaussian, sine or cosine distributions have also been considered by us. By these, we meant that the weights of $c's$ and $d's$ are distributed as in Gaussian, sine or cosine shape when plotted as functions of the angular momentum in the $z$ direction ($|m| \leq s$). For instance, when the Gaussian shape is considered, we make use of the (not-normalize) discrete Gaussian kernel defined by $T(m,\Delta) = c_m = e^{-\Delta} I_m(\Delta)$, where $\Delta$ is related to the Gaussian's standard deviation and $m$ takes the values between $\{-s, s\}$; $I_m(\Delta)$ denotes the modified Bessel functions of integer order, $m$. The above distributions gave us a quite similar cooling performance with similar success probability, slightly deviating from the flat distribution ---which explains why we explore the flat case as an illustrative example. With these parameters restrictions, the unnormalized density matrix is reduced to

\begin{eqnarray}
\nonumber \hat{\rho}(t)_\mathrm{post} &=& \frac{1}{(N + 1)^2}\sum_{\substack{m=-s,s\\ m'=-s',s'}} e^{i\lambda^2(t - \sin t)(m^2 - m'^2)}\\
&\otimes& \hat{D}[\lambda m \eta] e^{-i \hat{b}^\dagger \hat{b} t} \hat{\rho}(0)_\mathrm{NMO} e^{+i \hat{b}^\dagger \hat{b} t} \hat{D}[\lambda m' \eta]^\dagger.\label{eq:post_flat}
\end{eqnarray}

\begin{figure}
 \centering \includegraphics[width=\linewidth]{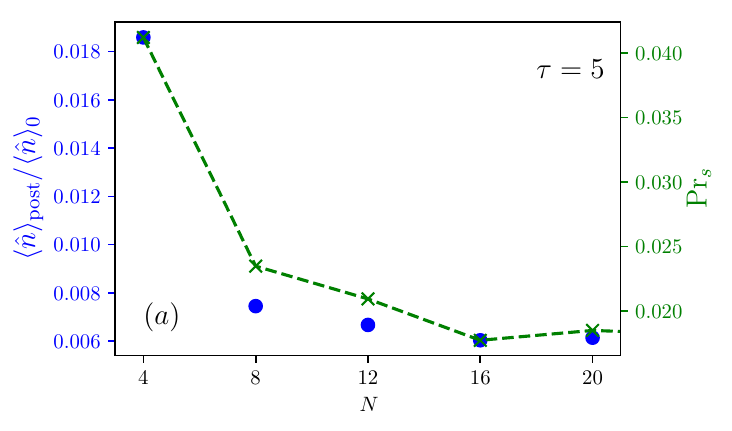}
 \centering \includegraphics[width=\linewidth]{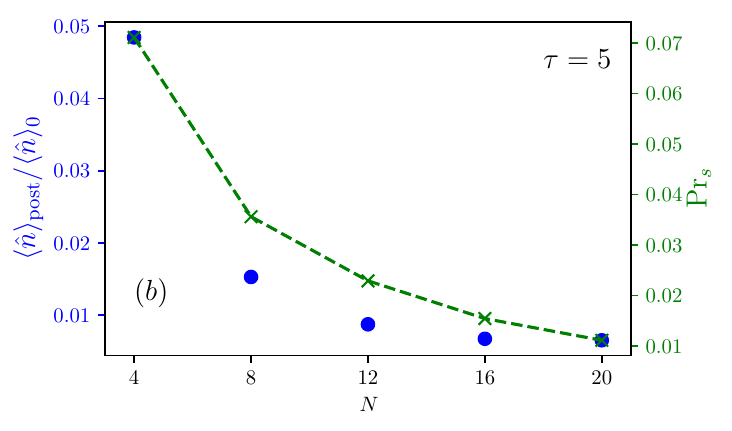} 
 \caption{Cooling rate (left y-axis, blue dots) and its related success probability (right y-axis, green crosses) as a function of the number of spins coupled to the NMO. We have fixed the total number of iterations $(\tau)$ as $\tau = 5$ (therefore, $t = \tau \times \pi /2$, corresponds to the total time of the protocol). In the top panel (a), we considered a flat distribution for the parameters $\{c's, d's\}$ as shown in Eq. (\ref{eq:post_flat}). In the bottom panel (b), we use the symmetric coherent spin state $\ket{\mathrm{CSS}}_+$ both for the preselection and postselection of the spins ---a state which proves to be the optimal case. In both cases, going beyond $N = 10$ spins does not improve the cooling performance substantially.} \label{fig:50}
\end{figure}

In Fig. \ref{fig:50}, we illustrate the benefits of considering the collective basis for the spin postselection. The advantage comes concerning the success probability and the total number of iterations. In the top panel (a), we present the cooling performance (left y-axis) and the success probability (right y-axis) as a function of the number of spins for a flat distribution [see Eq. (\ref{eq:post_flat})]. With this, we now intend to address whether there is a critical number of spins for which we can achieve an optimal balance between minimizing the cooling rate and maximizing the success probability as higher as the protocol can attain. From previous discussions (e.g., see Fig. \ref{fig3:fig3}), it is clear that as $\tau$ decreases we will require more spins to optimally achieve the NMO ground-state. Let us investigate the situation where we iterate our protocol for a maximum value of $\tau = 5$. From Fig. \ref{fig:50}-(a), it is seen that having more than $N = 10$ spins result into a misuse of resources as the differences between the cooling rate and the success probability become negligible when $N \geq 10$ (for fixed $\tau = 5$).

Nonetheless, despite the highly beneficial final state achieved so far (with $\langle \hat{n} \rangle_\mathrm{post} \approx 0.07$, for $N = 10$, and $\mathrm{Pr}_s \approx 2\%$), the election of a flat distribution for the pre- and post-selection parameters $\{c_i, d_j\}$ might seem purely of theoretical interests. In other words, we are unable to give a truly experimental preparation of this distribution, as we are blind between the flat distribution in Dicke basis and its transformed basis in the ``natural'' spin computational basis $\{0's, 1's\}$ or $\{\up 's, \down 's\}$. Certainly, this transformation matrix can be found straightforwardly in our simple case ($s = N/2$) as following

\begin{equation}
 \ket{\frac{N}{2}, m} = \frac{1}{\sqrt{\binom{N}{m + \frac{N}{2}}}}\hat{\mathcal{S}}\left[\up^{\otimes m + N/2} \otimes \down^{N/2 - m} \right],
\end{equation}

where, $\hat{\mathcal{S}}$ is the symmetrization operator.

As mentioned, even though we have found some other distributions of $c's$ and $d's$ where our protocol still work (e.g., a Gaussian distribution), we would like to focus our attention in states of experimental interests (or theoretical proposals to achieve them) (\cite{Agarwal2007, Wu2017, Hume2009, Toyoda2011, Zhou2011, Svetoslav2013, Neto2017} and references therein), such as the symmetric coherent spin state ($a = b = 1/\sqrt{2}$)

\begin{eqnarray}
\nonumber |\mathrm{CSS} \rangle_+ &=& \bigotimes_{i=1}^N(a\up_i + b\down_i)\\
\nonumber &=& \sum_{m=-N/2}^{N/2} \sqrt{\binom{N}{\frac{N}{2} + m}}a^{N/2 + m} b^{N/2 - m}\ket{\frac{N}{2}, m},\\
\end{eqnarray}

the antisymmetric coherent spin state $|\mathrm{CSS} \rangle_- = \otimes_i^N(a\up_i + b\down_i)$, with $a = 1/\sqrt{2}, b = - 1/\sqrt{2}$, the excited state $|1\rangle^{\otimes N} = \ket{N/2, N/2}$ (all spins up), ground state $|0\rangle^{\otimes N} = \ket{N/2, -N/2}$ (all spins down), the maximally-symmetric superradiant state $\ket{N/2, 0}$, the subradiant state $\ket{0 , 0}$, and finally, the Greenberger - Horne - Zeilinger (GHZ) state $1/\sqrt{2} \times (|0\rangle^{\otimes N} + |1\rangle^{\otimes N}) = 1/\sqrt{2}(\ket{N/2, N/2} + \ket{N/2, -N/2})$. Thanks to the Permutational Invariant Quantum Solver  (PIQS) \cite{Shammah2018} and the Quantum Toolbox in Python (QuTiP) \cite{Johansson2013}, the quantum dynamics considering the above spin states can be easily computed. Although we considered several combinations of the spin states, we have only found that the symmetric, antisymmetric and GHZ states are suitable to steer the mechanical state to its ground-state. From the previous discussed sections, both symmetric and antisymmetric coherent spin states are quite expected to assess such mechanical cooling, whereas GHZ states can also achieve near ground-state cooling as they simply can be written in the spin computational basis as following:

\begin{eqnarray}
 \hat{U}(t)\ket{\beta}\up^{\otimes N} &=& e^{i\lambda_\mathrm{eff}^2(t - \sin t)} e^{i\theta_\mathrm{eff}} \ket{\beta_{\uparrow,\mathrm{eff}}}\up^{\otimes N},\\
 \hat{U}(t)\ket{\beta}\down^{\otimes N} &=& e^{i\lambda_\mathrm{eff}^2(t - \sin t)} e^{-i\theta_\mathrm{eff}} \ket{\beta_{\downarrow,\mathrm{eff}}}\down^{\otimes N}
\end{eqnarray}

where $\beta_{\uparrow,\mathrm{eff}} = \beta e^{-it} + \lambda_\mathrm{eff} \eta$, $\beta_{\downarrow,\mathrm{eff}} = \beta e^{-it} - \lambda_\mathrm{eff} \eta$, $\theta_\mathrm{eff} = \lambda_\mathrm{eff} r \cos (\phi - \frac{t}{2})\sin \frac{t}{2}$, $\beta = r e^{i\phi}$ and $\lambda_\mathrm{eff} = \lambda N$. Hence, the (unnormalized) oscillator density matrix for a GHZ state postselected with itself is:

\begin{eqnarray}
 \nonumber \hat{\rho}(t) &=& \frac{1}{4\pi\bar{n}} \int d^2\beta e^{-\frac{|\beta|^2}{\bar{n}}}\Big(\ket{\beta_{\uparrow,\mathrm{eff}}}\bra{\beta_{\uparrow,\mathrm{eff}}} + \\
 \nonumber &&\ket{\beta_{\downarrow,\mathrm{eff}}}\bra{\beta_{\uparrow,\mathrm{eff}}} + [\ket{\beta_{\uparrow,\mathrm{eff}}}\bra{\beta_{\downarrow,\mathrm{eff}}}e^{2i\theta_\mathrm{eff}} +  h.c]\Big).\\
\end{eqnarray}

In Fig. \ref{fig:50}-(b) we summarize our findings for the optimal collective case, i.e., when the spin pre- and post-selection is performed using $\ket{\mathrm{CSS}}_+$. As in the previous case, the total time of the protocol consists of a total of five iterations. In this case, going beyond $N \geq 10$ spins is translated once again into a waste of spin resources. The cooling performance decreases as low as $\langle \hat{n} \rangle_\mathrm{post}/\langle \hat{n} \rangle_0 \approx 0.01$ with a success probability of $\mathrm{Pr}_s \approx 2\%$.

The conclusion so far can be stated as follows. Of all the states studied by us, the one it brings the mechanical object to its ground state more efficiently is the symmetric coherent spin state (from an experimental point of view), i.e., each spin state initialized individually in $\ket{+}$, for which the framework using the Dicke basis gives us a good advantage concerning computational time. Here, we would like to point out that, inhomogeneous coupling, namely different $\lambda_i$ can also be taken into account. However, the inhomogeneities forbids to represent the Hamiltonian in terms of the total angular momentum operators, i.e., $\hat{S}_z \neq \sum_{i=1}^N \hat{\sigma}_{z,i}$ (see Ref. \cite{Tsy2010}). Furthermore, it is also relevant to point out that, as we described, the postselection state is critical for our scheme. Here we have mainly focus in preselection and postselection of the spin in their same state, as in this manner we do not have any need of restarting the spin state after at each iteration. It is worthwhile to mention that, flat distributions of $c's$ and $d's$ parameters can also have a substantial impact in reducing the phonons on average; however, the experimental realization could be quite costly. With the above Dicke framework we can study the performance of our protocol in quantum open systems, which is the subject of the next Section.

\section{Open quantum case and experimental feasibility}\label{sec:master-equation}

An unavoidable fact in realistic scenarios is that all physical systems suffer from decoherence, i.e., detrimental effects due to the interaction of the system's relevant degrees of freedom with the reservoir. This is of special interests for any cooling scheme, as we need to cool down the NMO by competing with the thermalization due to the environment. To simulate this scenario, let us first consider the open quantum case in presence of both mechanical damping in a reservoir with $\bar{n}$ initial phonon accupation number, and local spin relaxation and also including local pure dephasing terms. Therefore, by considering an initial NMO embedded in a reservoir with $\bar{n} \equiv \langle \hat{n} \rangle_0 \approx 10$ thermal phonon number on average, we then commence to evolve (at each iteration) the spin-mechanical dynamics from a non-equilibrium state, i.e., where $\bar{n} > \langle \hat{n} \rangle_\mathrm{post}$. The above open quantum case can be modeled by the usual master equation in Markov-Born approximation as follows:

\begin{eqnarray}
\nonumber \frac{d\hat{\rho}}{dt} &=& -i[\hat{H}_\mathrm{int},\hat{\rho}] + \gamma(1+\bar{n})\mathcal{L}[\hat{b}] + \gamma\bar{n}\mathcal{L}[\hat{b}^{\dagger}] \\
\nonumber &+& \sum_{i=1}^{N} \Gamma(1+\bar{n})\mathcal{L}[\hat{\sigma}^-_i] + \Gamma\bar{n}\mathcal{L}[\hat{\sigma}^+_i] + \frac{\gamma_\phi}{2}\mathcal{L}[\hat{\sigma}_{z,i}]\\
\label{master2}
\end{eqnarray}

where $\hat{H}_\mathrm{int} = \hat{b}^\dag\hat{b} - \sum_{i=1}^N \lambda_i \hat{\sigma}_{z,i}(\hat{b}^\dag + \hat{b})$  and
\begin{equation}
  \mathcal{L}\left[\hat{O}\right] = \frac{1}{2}\left(
  2\hat{O}\hat{\rho}\hat{O}^{\dagger}-
  \hat{\rho}\hat{O}^{\dagger}\hat{O}-\hat{O}^{\dagger}\hat{O}\hat{\rho}\right)
\end{equation}

corresponds to the Lindblad term. Furthermore, the scaled (by the mechanical frequency $\omega_m$) quantities $\{\gamma, \Gamma, \gamma_\phi\}$ are the local mechanical damping, local spin relaxation, and the local spin pure dephasing rates, respectively. To estimate up to which values our scheme can be accommodated, we consider the simulation for $N=1$ and $N=4$ spins. By a numerical simulation we found that the scheme does not suffer heavily regarding spin decoherence, for instance, $\{\Gamma, \gamma_\phi\} \lesssim \{10^{-3},10^{-2}\}$ \cite{Bar-Gill2013} will not produce any strong effect during the iterations. On the other hand, we considered the simulation with mechanical states of (scaled) quality factors as low as (Q=$\gamma^{-1}$) $Q \sim 10^{3}$, nowadays, $Q > 10^3$ can be achieved experimentally \cite{Aspelmeyer2014}.

Nonetheless, more appropriate for this type of open quantum dynamics is to consider both local as well as collective losses. The associated master equation for this general case is:

\begin{eqnarray}
\nonumber \frac{d\hat{\rho}}{dt} &=& -i[\hat{H}_\mathrm{int},\hat{\rho}] + \gamma(1+\bar{n})\mathcal{L}[\hat{b}] + \gamma\bar{n}\mathcal{L}[\hat{b}^{\dagger}] \\
\nonumber &+& \sum_{i=1}^{N} \Gamma(1+\bar{n})\mathcal{L}[\hat{\sigma}^-_i] + \Gamma\bar{n}\mathcal{L}[\hat{\sigma}^+_i] + \frac{\gamma_\phi}{2}\mathcal{L}[\hat{\sigma}_{z,i}]\\
&+& \Gamma_\Downarrow \mathcal{L}[\hat{S}^-] + \Gamma_\Uparrow \mathcal{L}[\hat{S}^+] + \frac{\gamma_\Phi}{2}\mathcal{L}[\hat{S}_{z}]
\label{master_collective}
\end{eqnarray}

where the corresponding collective phenomena are described by the scaled quantities $\{\Gamma_\Downarrow, \Gamma_\Uparrow, \gamma_\Phi\}$, being the collective decay (typical of superradiant decay), the collective pumping (which in this case can be understood as incoherent pumping from the thermal reservoir), and collective dephasing, respectively.

\begin{figure}
 \centering \includegraphics[width=\linewidth]{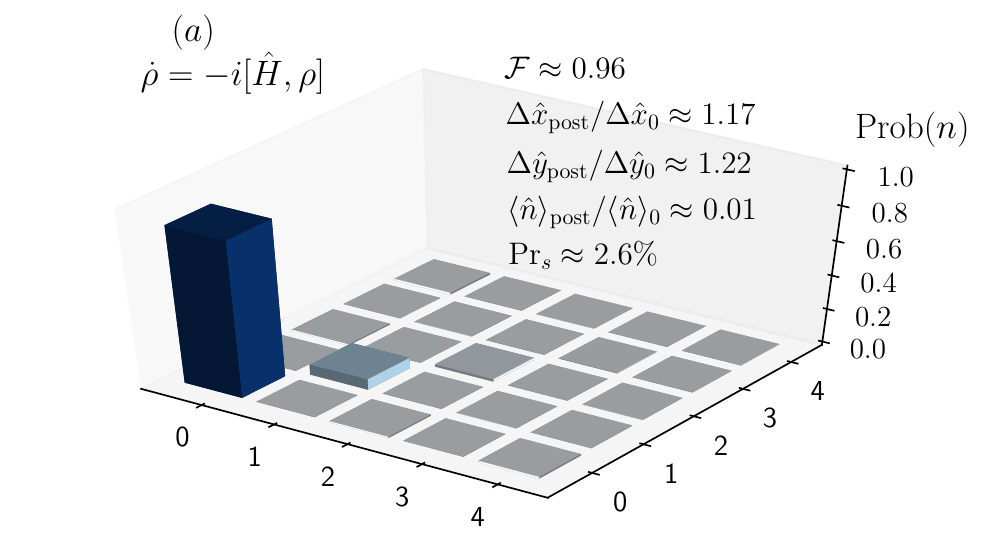}
 \centering \includegraphics[width=\linewidth]{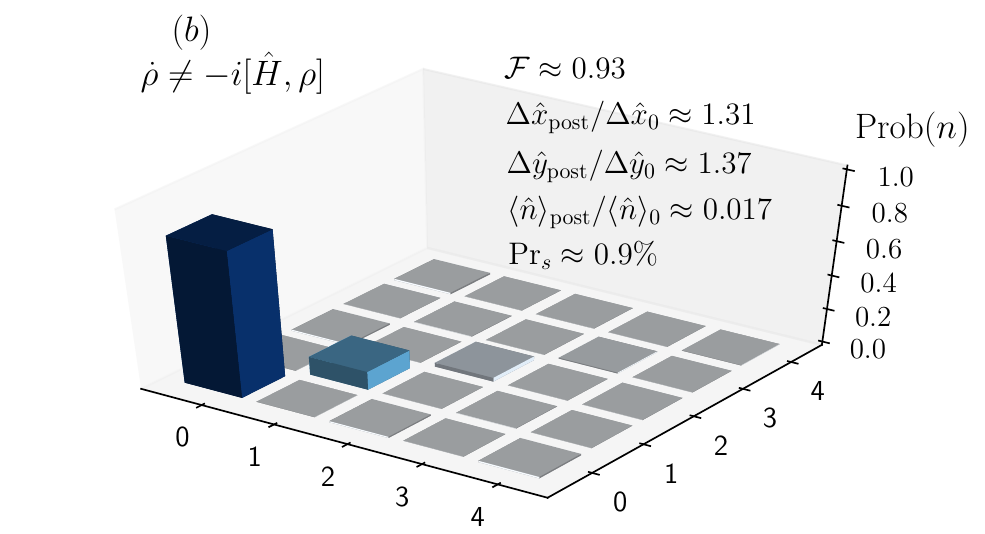} 
 \caption{Phonon occupation probability [$\mathrm{Prob}(n)$] for the postselected mechanical state. In (a) we solved the unitary evolution and (b) the open quantum case, where we have considered both local and collective channels of dissipation $\{\gamma = \Gamma, \gamma_\phi = \Gamma_\Downarrow = \Gamma_\Uparrow = \gamma_\Phi, \langle \hat{n} \rangle_0\} = \{10^{-3}, 10^{-2}, 10\}$. As before, we simulated the protocol up to five iterations ($t = \tau \times \pi/2 = 5 \times \pi/2$) when $N = 10$ spins are coupled to the NMO. Preparation and postselection of the spins have been performed in the state $\ket{\mathrm{CSS}}_+$.} \label{fig:histograms}
\end{figure}

In Fig. \ref{fig:histograms} we contrast the final NMO state using our protocol in the absence of any source of decoherence $(a)$ with the one including several channels of dissipation $(b)$. To observe the robustness of our scheme in the presence of decoherence, let us first consider the unitary evolution and explore up to which values our protocol can be accommodated. In the top panel, we depicted the phonon occupation probability $\bra{n} \hat{\rho}_m \ket{n}$ for the final (postselected) mechanical state after five iterations when $N = 10$ spins started from the $\ket{\mathrm{CSS}}_+$ and they are postselected accordingly; the NMO was initialized, as throughout this work, with $\bar{n} = 10$ phonons on average. As seen from the figure, at the end of the protocol, the phonons on average have been decreased as low as $\langle \hat{n} \rangle_\mathrm{post}/\langle \hat{n} \rangle_0 \approx 0.01$ with a success probability of $\mathrm{Pr}_s \approx 2.6\%$. Additionally, we have included the ratio between the mechanical variances after the protocol takes place with the mechanical variances of the pure ground-state $\ket{0}_m$, where $\Delta \hat{x}_\mathrm{post}/\Delta \hat{x}_0 \approx 1.17, \Delta \hat{y}_\mathrm{post}/\Delta \hat{y}_0 \approx 1.22 $. It is then readily observed that, the final NMO state truly resembles the ground-state of the oscillator, with a fidelity $\sqrt{ \bra{0}_m \hat{\rho}(t)_\mathrm{post} \ket{0}_m} \approx 0.96$. In the bottom panel (b), we solve the dynamics including several dissipation channels with both local and collectives Lindbladians, where we considered $\{\gamma = \Gamma, \gamma_\phi = \Gamma_\Downarrow = \Gamma_\Uparrow = \gamma_\Phi, \langle \hat{n} \rangle_0\} = \{10^{-3}, 10^{-2}, 10\}$ (see next paragraph regarding experimental feasibility). This result shows that even in the case of decoherence [as general as the one described by the master equation in Eq. (\ref{master_collective})] the protocol is positively robust. We can understand this robustness regarding the total number of iterations and, especially, the total protocol time $t = 5 \times \pi/2$ (approximately two NMO whole cycles), as it has been highly reduced because of the inclusion of several spins coupled to the NMO. Therefore, at each spin postselection both local and collective channels of dissipation do not have enough time to thermalize the collapsed NMO. In this last open quantum scenario, $\langle \hat{n} \rangle_\mathrm{post}/\langle \hat{n} \rangle_0 \approx 0.017$ with a success probability of $\mathrm{Pr}_s \approx 0.9\%$, with an approximate fidelity of $\approx 0.93$. And, the ratio of mechanical variances are $\Delta \hat{x}_\mathrm{post}/\Delta \hat{x}_0 \approx 1.31, \Delta \hat{y}_\mathrm{post}/\Delta \hat{y}_0 \approx 1.37$.

Although we have not given any specific physical system for our cooling scheme so far, we believe that a feasible system could be the one involving micro- or nano-cantilevers coupled to nitrogen-vacancy (NV) centers in diamond \cite{Rabl2009, Ma2016, Bhaktavatsala-2016, Treutlein2014}. As also stated in Ref. \cite{Bhaktavatsala-2016}, solid-state spins are robust regarding long coherence \cite{Waldherr2014, Bhaktavatsala-2016} as well as long relaxation times $T_1$ that reaches few milliseconds at room temperature \cite{Jarmola2012}. For instance, it is well known that the electron spin coherence time measured by a Ramsey experiment is around $T_2^* = 1.35 \mu s$ \cite{Blok2014}. At first sight, this time might limit the proper implementation of our protocol. However, increments of the coherence time have been observed in spin echo measurements $T_2 = 395 \mu s$ \cite{Zaiser2016}  or $T_2 = 1.8 ms$ in an isotopically engineered diamond \cite{2009NatMa}. Moreover, they can also be easily initialized and readout with fidelities exceeding 98\% \cite{Robledo2011, Doherty2013}. On the other hand, the spin-mechanical coupling can be obtained through magnetic coupling, where the scaled coupling parameter reads as $\hbar \lambda \approx \mu_B \partial B / \partial z \sqrt{\hbar / 2m\omega_m^3}$ \cite{Treutlein2014, Rabl2009}. A typical set of values are $\mu_B \sim 10^{-23}$ J/T (Bohr magneton), mass $m \sim 10^{-14}$ kg, mechanical frequency $\omega_m \sim 10^6$ Hz, and magnetic gradient between $10^4$ T/m $< \partial B / \partial z < 10^7$ T/m, thus enabling the needed operational regime $10^{-4} < \lambda < 10^{-1}$ \cite{Rabl2009, Treutlein2014}.

Finally, preparation, measurement, and control of collective spin states in experiments related to quantum optics, solid state physics, etc. \cite{2016arXiv160901609P}, has become nowadays, in a plausible technic. For example, the pioneer experiments as \cite{Hald1999, Kuzmich2000, Julsgaard2001} are well-known protocols for realizing squeezing, teleportation, and entanglement of macroscopic atomic samples with engineered collective spin states \cite{Bohnet2014}. Recently, new experimental proposals consider the hybrid systems, where the main actors could be the collective spin-like states (atoms, superconducting qubits, NV centers) coupled, e.g., to superconducting resonators \cite{Kakuyanagi2016, 2017arXiv171204357C}, or mechanical elements \cite{Ceban2017, MacQuarrie2017}. Particularly in Ref. \cite{MacQuarrie2017}, the authors propose an experimental protocol to cool down a high-Q mechanical resonator from room temperature. This goal is achieved by coupling collectively an high-density ensemble of NV-centre spins to a mechanical oscillator via an (excited state) spin-strain mechanism.

\section{Final remarks}\label{sec:final-remarks}

In summary, we have explored different alternatives to cool-down a nanomechanical oscillator (NMO) to its ground-state when $N$ spins are coupled independently to the NMO position. Our probabilistic protocol relies on successive spin postselections iterations conducted at each NMO quarter of its cycle ($t = \pi/2$). For the simplest case of having only a single spin coupled to the NMO, we have found an optimal interaction time and spin-mechanical coupling, a set of values that must happen to reduce the NMO position and momentum variances simultaneously ---other coupling strength at some other time lead the NMO to mechanical squeezing. We would like to mention that the single spin case does not depart much as the work by D. D. B. Rao \textit{et al.}. Nevertheless, our primary aim was to investigate the cooling process in the presence of $N \neq 1$ spins. Moreover, as the process is intrinsically iterative, one would require to succeed at any stage to efficiently cool-down the NMO; otherwise, a failure spin postselection will drive the NMO to an intricated quantum state, and therefore a full reinitialization of the scheme is necessary. When $N$ spins are coupled to the NMO, the inquiry on how to optimally postselect the spins become highly nontrivial. For this reason, we consider three main cases regarding the conditioned spin measurement, namely i) individual postselection of the spins, ii) in correlated basis, and iii) of making use of collective operators. The latter one is typically used when $N$ independent atomic emitters are coupled identically to a quantized electromagnetic field ---for example, being trapped or located them in the antinodes of a standing light wave. Here, we found that the total number of iterations needed to bring the NMO near to its ground state is highly reduced, where also the final state is achieved with a non-negligible success probability. The reduction of the whole protocol time shows to be vastly beneficial when the relevant system is in contact with a thermal reservoir at $T \neq 0$. To model this situation we consider a general master equation shown in Eq. (\ref{master_collective}), where we included local and collective channels of dissipation. We found that even in this general decoherence picture our scheme can be accommodated up to feasible values (all scaled by the NMO frequency) $\{\gamma = \Gamma, \gamma_\phi = \Gamma_\Downarrow = \Gamma_\Uparrow = \gamma_\Phi, \langle \hat{n} \rangle_0\} = \{10^{-3}, 10^{-2}, 10\}$, where $\langle \hat{n} \rangle_\mathrm{post}/\langle \hat{n} \rangle_0 \approx 0.017$, success probability of $\mathrm{Pr}_s \approx 0.9\%$, an approximated fidelity of $0.93$, and mechanical variances $\Delta \hat{x}_\mathrm{post}/\Delta \hat{x}_0 \approx 1.31, \Delta \hat{y}_\mathrm{post}/\Delta \hat{y}_0 \approx 1.37$.

\section*{ACKNOWLEDGMENTS}
V.M. and R.C acknowledge the financial support of the projects Fondecyt Postdoctorado $\#$3160700 and $\#$3160154, respectively. M.O. and V.E. acknowledge the financial support of the project Fondecyt $\#$1180175. V.M. would like to thank Nathan Shammah for his valuable suggestion on using the Permutational Invariant Quantum Solver (PIQS), an open source library in Python; we were highly benefited from his feedback.

\bibliography{Cooling}

\end{document}